\documentstyle[epsf,aps]{revtex}
\newcommand{\bea}{\begin{eqnarray}}
\newcommand{\eea}{\end{eqnarray}}        
\newcommand{\be}{\begin{equation}}
\newcommand{\ee}{\end{equation}}
\newcommand{\bt}{\begin{tabular}}      
\newcommand{\et}{\end{tabular}}
\newcommand{\Tr}{{\rm Tr}}
\newcommand{\no}{\nonumber}
\newcommand{\ovl}{\overline}

\newcommand{\ds}{\displaystyle}
\newcommand{\pa}{\partial}
\begin{document}
\title{Chiral Lagrangian for strange hadronic matter}
\author{ P.~Papazoglou$^{a}$,
S. ~Schramm$^{b}$, 
J. ~Schaffner-Bielich$^{c}$,
H. ~St\"ocker$^{a}$,
W. ~Greiner$^{a}$}
  \address{
        $^a$Institut f\"ur Theoretische Physik, 
        $^{~}$Postfach 11 19 32, D-60054 Frankfurt am Main, Germany\\
        $^b$GSI Darmstadt, Postfach 11 05 52, D-64220 Darmstadt, Germany\\
        $^c$Nuclear Science Division, LBNL, 1 Cyclotron Road, Berkeley, 
CA94720
} 
\date{\today}
\maketitle
\begin{abstract}
A generalized Lagrangian for the description of hadronic matter based 
on the linear $SU(3)_L \times SU(3)_R$ $\sigma$-model is proposed. 
Besides the baryon octet, the spin-0 and spin-1 nonets, a gluon condensate 
associated with broken scale invariance is incorporated. 
The observed values for the vacuum masses of the baryons and mesons
are reproduced. In mean-field 
approximation, vector and scalar interactions yield a 
saturating nuclear equation of state. We discuss the difficulties and 
possibilities to construct a chiral invariant baryon-meson interaction that 
leads to a realistic equation of state. It is found that a 
coupling of the strange condensate to nucleons is needed to describe 
the hyperon potentials correctly. The effective baryon masses and the 
appearance of an abnormal phase of nearly massless nucleons at high 
densities are examined. A nonlinear realization of chiral symmetry
is considered, to retain a Yukawa-type baryon-meson interaction and 
to establish a connection to the Walecka-model.

\end{abstract}
\pacs{PACS numbers:12.39.Fe, 11.30.-j, 21.30.Fe}

\section{Introduction}
Recently, nuclear physicists have given new attention to the  
general principles of chiral symmetry and broken scale invariance 
at finite densities. 
The underlying theory of strong interactions, QCD, is presently 
not solvable in the nonperturbative regime of low energies. However, 
QCD constraints may be imposed on effective theories for nuclear physics 
through symmetries, which determine largely how the hadrons should interact 
with each other. In this spirit, models with  
$SU(2)_L \times SU(2)_R$ symmetry and scale invariance were 
applied to nuclear matter at zero and finite temperature 
and to finite nuclei\cite{heid94,cart95,furn95,mish93,paper1}. The 
success of these models  established 
the applicability of this approach  to  relativistic 
nuclear many-body physics.\\  
A simultaneous and self-consistent description of strange and nonstrange 
particles in baryonic matter is of particular interest, since many 
questions in heavy-ion physics and astrophysics are related to the 
effect of strangeness in matter: The possible large strangeness content 
of the nucleon indicates the importance of taking strangeness into account 
for a deeper understanding of nuclear matter and nuclei\cite{strangenuc}. 
When extrapolating to hadronic systems with a large amount of strangeness, 
new phenomena as negatively charged multistrange objects may occur\cite{scha93}. 
The possible onset of kaon condensation at high baryon densities in heavy-ion 
collisions and the interior of neutron stars provides another motivation for 
studying models which include the strange degree of freedom. \\
Hadrons can be classified in multiplets with (broken) $SU(3)_V$ symmetry, 
in which they have (almost) degenerate masses \cite{Eight64}. 
If there is one limit to the strong interactions, in which $SU(3)_V$ is 
exact, and another one, in which $SU(2)_V \times SU(2)_A$ symmetry holds, then 
there must be a joint limit in which $SU(3)_V$ is exact {\it and} all the 
axial-vector currents are conserved. In this limit the $\pi$, $K$, 
and $\eta$-particles are Goldstone bosons 
and we are led to a Lagrangian invariant under $SU(3)_L \times SU(3)_R$.\\
The linear $\sigma$-model as a specific realization of $SU(3)_L \times SU(3)_R$ 
symmetry was extensively studied  
in free space\cite{levy67,gasi69,sche71}. The spin-0 mass spectrum and 
meson-nucleon scattering are satisfactorily described within this approach.\\
In this paper we investigate the applicability of chiral SU(3) symmetry 
to describe nuclear matter properties by constructing a chiral Lagrangian 
for hadronic matter including 
strange particles. To reproduce the binding energy of nuclear matter at 
saturation density
 $\rho=0.15 \rm ~fm^{-3}$ with a reasonable value for the compression modulus, 
an octet of vector mesons with axial mesons as chiral partners is included.\\
The work is based on studies of nuclear matter 
with the $SU(2)_L\times SU(2)_R$ linear $\sigma$-model 
\cite{heid94,cart95,furn95,mish93,paper1}. There, a logarithmic 
potential involving the dilaton field $\chi$ 
introduced in \cite{sche80} to mimick the trace 
anomaly of QCD plays an essential role. It eliminates unphysical 
bifurcations encountered in the linear $\sigma$-model when applied 
to describe nuclear matter properties\cite{kerman}. 
A similar concept is adopted here, too.   
However, there are some important difference to the SU(2) case:
the extension to a chiral SU(3) symmetric model is nontrivial,
because ---in contrast to the nucleon doublet--- 
the baryon octet cannot be assigned to a fundamental representation. 
Because of this, difficulties in describing 
the baryon masses and the hyperon potentials simultaneously arise.
Furthermore, one needs to reproduce the experimentally well known masses of the
baryon octet and the meson nonets. This leads necessarily 
to the inclusion of cubic and quartic self-interactions of spin-0 mesons, which 
were eliminated in \cite{heid94} to improve their results for 
nuclear matter and nuclei.\\
The specific form of baryon--meson interaction is crucial for 
the properties of (hyper--)nuclear matter. The (relativistic) potentials for 
nucleons and hyperons following from this model depend strongly on 
the coupling constants of hyperons to vector and scalar mesons. Since these 
are constrained by chiral symmetry, it is of interest, whether or not 
the hyperon 
potentials are described correctly within this approach. Furthermore, the 
way hyperons are treated has important consequences for the stability 
of multistrange hypernuclear systems and for the mass of neutron stars. 
Therefore, different forms of coupling baryons to spin-0 mesons 
and their 
influence to the hyperon potentials are examined.\\
The paper is organized as follows: In the first part we review the chiral 
transformations of
mesons and baryons and their assignment to representations.
Then, the Lagrangian in its general form is presented and discussed.
The next part is devoted to the approximation scheme used.
The results include the investigation of the equation of state for nuclear matter, 
the hyperon potentials, the effective baryon masses, and a discussion 
whether a chiral phase transition occurs at high densities. 
As an outlook, a nonlinear realization of chiral symmetry is examined that 
is a convenient way to include heavy hadrons.
\section{Theory}
\label{theory}
%
%
The $\sigma$-model has been used extensively in exploring the implications
of chiral symmetry in
low-energy hadron dynamics. Most of these investigations have employed the
SU(2) model with mesons and nucleons and the SU(3) $\sigma$-model 
with mesons only.
In this section we will discuss the transformation properties of 
spin-0 and spin-1 mesons as well as those 
of the baryons as the constituents of our effective theory. This 
implies the choice of the proper representation under which the particles 
transform.
\subsection{Representations}
\label{repr} 
The representations of the hadrons result from
 the direct product of the quark representations, however in the Lagrangian 
there will be no explicit reference to quarks. For our purpose, they are 
used as guidance.\\ 
In the chiral limit, the quarks have to be massless. 
Therefore, it is sufficient to consider the 2-component spinors
\bea
    q_L &=& \frac{1}{2} (1-\gamma_5)q \qquad \sim \qquad (3,0)\\ \no 
    q_R &=& \frac{1}{2} (1+\gamma_5)q \qquad \sim \qquad (0,3) .
\eea       
Since the quarks are massless, the chirality of the spinors is linked to 
their spin. On the right-hand side, the quark representations are symbolized 
by the number of flavors, placed left (right) 
for the left (right) subspace of $SU(3)_L \times SU(3)_R$. 
\subsubsection{Mesons}
\label{mesons}
The mesons as a bound system of a quark and antiquark 
correspond to the bilinear form $\ovl{q} {\cal O} q$ where the 
12$\times$12 matrix ${\cal O}$ is the direct product of the 
4$\times$4 Dirac matrices and the 3$\times$3 Gell-Mann matrices 
(${\cal O} = \Gamma \otimes \lambda$). For the discussion of the 
representations we will first suppress the explicit reference to the 
Gell-Mann matrix $\lambda$. \\
Consider first the spin-0 mesons. If we suppose that they are 
$s$-wave bound states, then the only
spinless objects we can form are 
\be
 \ovl{q}_R   q_L  \qquad   \ovl{q}_L q_R .  
\ee
The combinations $\ovl{q}_L q_L$ and $\ovl{q}_R q_R$ vanish, because the 
left and right subspaces are orthogonal to each other.  
The resulting representation is (3,3$^{\ast}$) and
(3$^{\ast}$,3), respectively (the antiparticles belong to the 
conjugate representation). We are thus led to consider nonets of
pseudoscalar and scalar particles.\\  
For the vector mesons, we have to construct vectorial 
quantities from 
$q_L$ and $q_R$. Again, if we assume that s-wave bound states are involved, 
the only vectors which can be formed are
\be
 \ovl{q}_L \gamma_{\mu} q_L  \qquad \ovl{q}_R \gamma_{\mu} q_R .  
\ee
This suggests assigning the vector and axial vector mesons to the
representation ($3 \times 3^{\ast}$,0) $\oplus$ (0,$3 \times 3^{\ast}$) = 
 (8,1) $\oplus$ (1,8), coinciding with the tensor
properties of the currents conserved in the SU(3) $\times$ SU(3) 
limit \cite{marsh65,gell64}.
\subsubsection{Baryons}
The discussion of baryons differs from that of the
mesons in that the construction of baryon multiplets from the basic fields
$q_L$ and $q_R$ is not unique. The reason is that a left- {\it or} right-
handed
quark can be added to the spin-0 diquark of one subspace. Consequently,
the baryons can be assigned to the representation $(3,3^{\ast})$ and
($3^{\ast}$,3) or (8,1) and (1,8), respectively. For an explicit
construction in terms of quark fields see \cite{ioff81,chri86}.
\subsection{Transformations}
Once the chiral-transformation properties of the elementary spinors
are known it is straightforward to derive the corresponding transformation
properties of the composite fields.
An arbitrary element of SU(3) $\times$ SU(3) can be written as
\bea
\label{lee9}
G(\alpha, \beta) &=& e^{\ds -[i \alpha_a Q^a +i \beta_a Q^{5a}]}  \\ \nonumber
&=& e^{\ds -i( \alpha- \beta)\cdot Q_{L}} \cdot e^{\ds -i( \alpha+ \beta)\cdot Q_{R}}
\equiv L \cdot R  .                      
\eea
where $\alpha$ and $\beta$ are eight-component vectors, and $Q^a$, $Q^{5 a}$ are 
the vector and axial generators of SU(3), respectively. 
The spinors $q_L$ of SU(3)$_L$ generated by $Q_L=(Q-Q^5)/2$ and $q_R$ of $SU(3)_R$ 
generated by $Q_R=(Q+Q^5)/2$ transform as
\be                                          
\label{spinor}
%
q_{j}' = L^k_j q_{k}  \qquad   q_{\ovl{j}}' =R^{\ovl{k}}_{\ovl{j}} q_{\ovl{k}}, 
\ee    
where we adopted the tensor notation.
Here, the (un-)bared indices belong to the (left) right subspace.  
The complex conjugate spinors transform as 
\be
\label{konspinor}
q'^{j} = (L^k_j)^{\ast} q^{k}  \qquad   q'^{\ovl{j}} 
=(R^{\ovl{k}}_{\ovl{j}})^{\ast} q^{\ovl{k}}. 
\ee 
Knowing the representation of the mesonic and baryonic fields,
it is straightforward to derive their transformation properties. They are
summarized in table \ref{chiraltrafo}, where we 
expressed the meson and baryon fields conveniently in a basis of 3 $\times 3$ 
Gell-Mann 
matrices. For example, the spin-0 mesons may be written in the compact form 
\begin{eqnarray}
\frac{1}{2}\sum_{a=0}^8 (\ovl{q}_R \lambda_a q_L+\ovl{q}_R \lambda_a 
\gamma_5 q_L)\lambda_a 
\equiv \frac{1}{\sqrt{2}}\sum_{a=0}^8 (\sigma_a +i \pi_a)\lambda_a 
= \Sigma+i \Pi &=& M  \\
\frac{1}{2}\sum_{a=0}^8 (\ovl{q}_L \lambda_a q_R+\ovl{q}_L \lambda_a \gamma_5 q_R) 
\lambda_a                        
\equiv \frac{1}{\sqrt{2}}\sum_{a=0}^8 (\sigma_a-i \pi_a)\lambda_a
= \Sigma-i\Pi &=&M^{\dagger} , 
\end{eqnarray}
where $\sigma_a=\ovl{q}\lambda_aq/\sqrt{2}$ (and similar for the $\pi_a$-fields) including 
the diagonal matrix $\lambda_0=\sqrt{\frac{2}{3}} \, 1{\hspace{-2.5mm}1}$.
The first and second row are connected by the parity transformation, which 
transforms left-handed quarks to right handed ones. This is achieved 
in the matrix formulation by taking the adjoint. Therefore, since scalar and 
pseudoscalar particles have opposite parity, an imaginary unit $i$ is attached 
to the pseudoscalar matrix $\Pi$. 
\subsection{Lagrangian formulation}
\subsubsection{Baryon-meson interaction}
\label{bmint}
\begin{figure}
\epsfbox{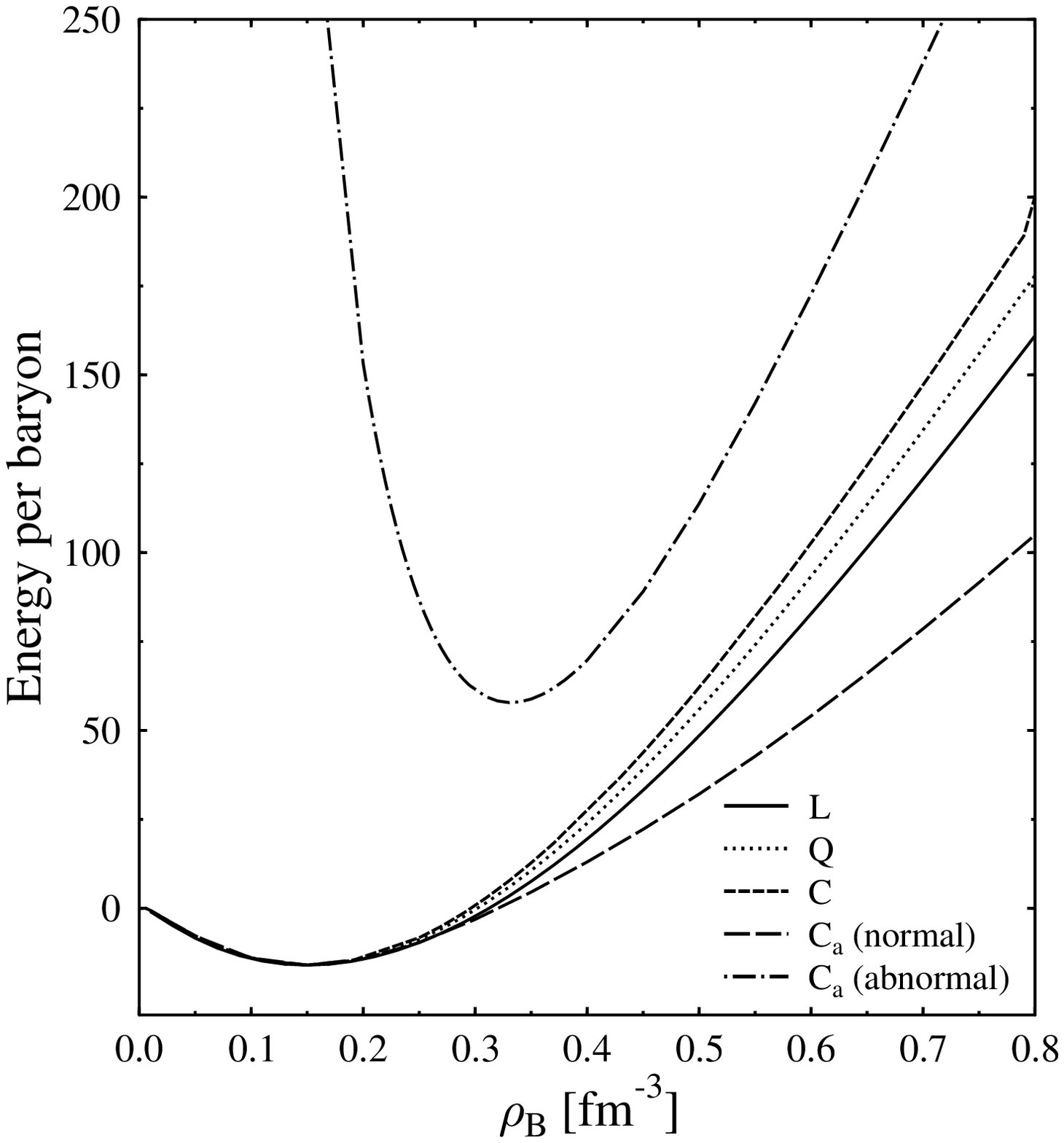}
\caption{Binding energy versus baryon density $\rho_B$ for the 
linear (L), quadratic (Q), and cubic (C) baryon-spin-0 meson interaction 
(see table \ref{parameter}).}
\label{energie2} 
\end{figure}

\begin{figure}
\epsfbox{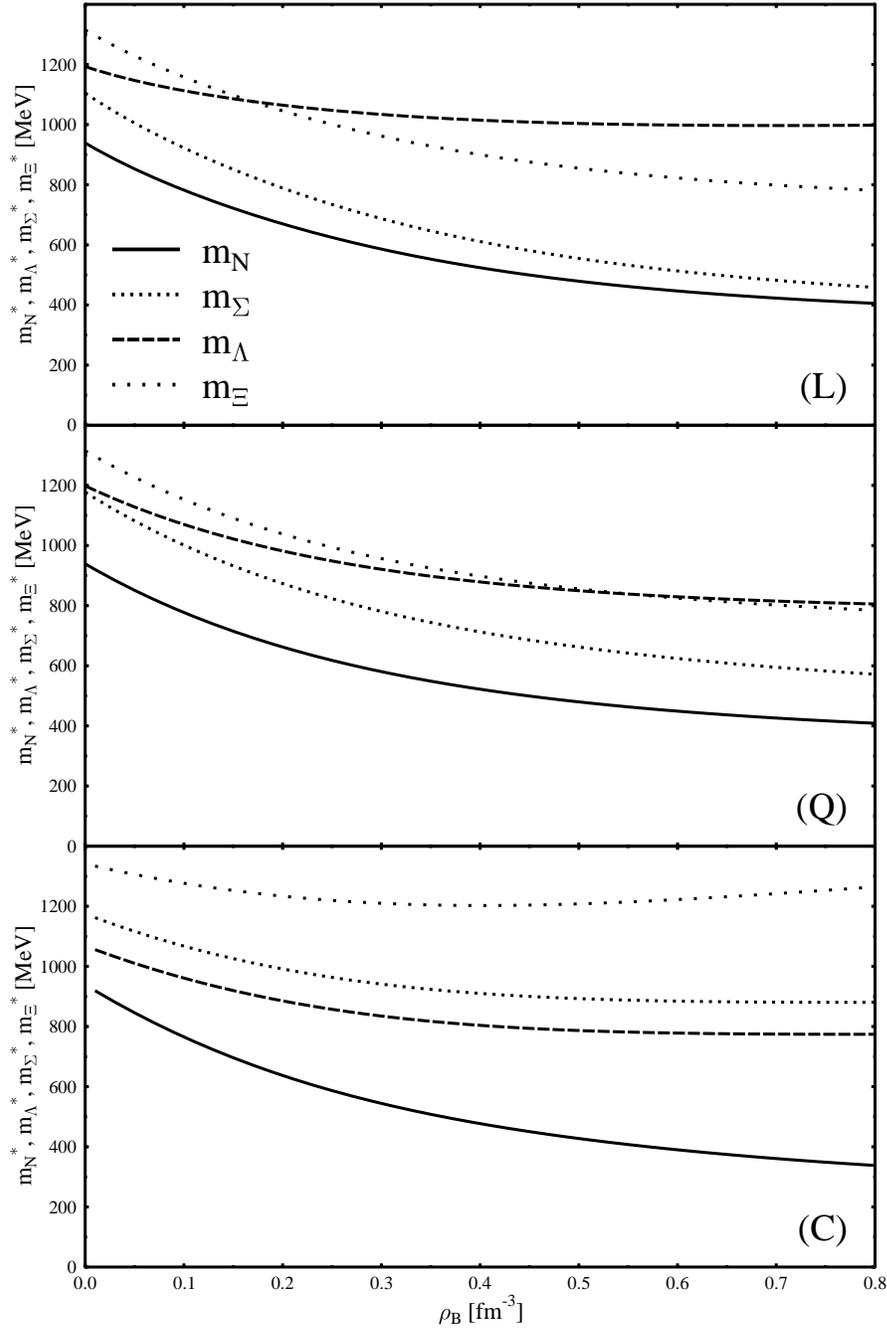}
\caption{\label{baryon} Baryon masses as a function of the baryon density 
$\rho_B$ for the 
linear (L), quadratic (Q), and cubic (C) baryon-spin-0 meson interaction.}
\end{figure}

When generalizing from SU(2) to SU(3), complications arise from the baryon-meson
sector, since not only the nucleon mass but the masses of the 
whole baryon multiplet are generated spontaneously by the vacuum 
expectation values (VEV) of only {\it two} meson condensates: of the 18 meson fields 
$\sigma_a$ and $\pi_a$ only the VEV of the components proportional to $\lambda_0$ and the 
hypercharge $Y \sim \lambda_8$ are nonvanishing, and the vacuum expectation 
value $\langle M \rangle$ reduces to: 
\begin{displaymath}
\langle M \rangle=\frac {1}{\sqrt{2}}(\sigma_0 \lambda_0+\sigma_8 \lambda_8)
\equiv \mbox{diag } (\frac{\sigma}{\sqrt{2}} \,, \frac{\sigma}{\sqrt{2}} \,, 
\zeta ) , 
\end{displaymath}
in order to preserve parity invariance and 
assuming, for simplicity, $SU(2)$ symmetry\footnote{This implies that 
isopin breaking effects will not occur, i.e., all hadrons of the 
same isospin multiplet will have identical masses.} of the vacuum. 
The quark content of these fields is 
$\sigma \sim \langle \ovl{u}u+\ovl{d}d \rangle$ and 
$\zeta \sim \langle \ovl{s}s \rangle$.
To see explicitly how these condensates generate the baryon masses, let us 
consider the simplest ansatz for the baryon-meson interaction, namely the 
Yukawa-type coupling:
\be
\label{bm2}
{\cal L}_{BM}^{(0)} = b_0\left( \varepsilon_{ \ovl{a} \ovl{b} \ovl{c} } 
\varepsilon^{def}
             (\ovl{\Psi}_L)^{\ovl{a}}_d M^{\ovl{b}}_e (\Psi_R)^{\ovl{c}}_f
          +  \varepsilon_{ abc } \varepsilon^{ \ovl{d} \ovl{e} \ovl{f} }
             (\ovl{\Psi}_R)^a_{\ovl{d}} M^b_{\ovl{e}} (\Psi_L)^c_{\ovl{f}}\right) \, . 
\ee
The indices are contracted appropriately to yield a chirally invariant term. 
Note that the chirally invariant {\it linear} baryon-meson 
interaction is only possible in the baryon
representation (3,$3^{\ast}$) and ($3^{\ast}$,3) and it is unique 
(since the product $3 \times 3  \times 3=1+ ...$ leads only to one singlet). 
Furthermore, the resulting coupling constants are given by the 
symmetric (d-type) structure constants of SU(3). The reason for this is that three 
spinors can only be coupled to a singlet by antisymmetrizing them. 
Since this has to be done in the left 
and right space, respectively, the resulting coupling will be a symmetric one. 
\\
Using the decomposition of the baryon matrix 
$\Psi=\frac{1}{\sqrt{2}} \sum_{k=0}^{8} \psi_k \lambda_k$  
by means of the projection operators $(1 \pm \gamma_5)/2$,
\be
\label{barlr}
(\Psi_L)^{\ovl{a}}_b = \frac{1 - \gamma_5}{2} \Psi^a_b \, , \qquad  
(\Psi_R)^a_{\ovl{b}} = \frac{1 + \gamma_5}{2} \Psi^a_b\, ,
\ee
one arrives at 
\bea
\label{bm0phys}
{\cal L}_{BM}^{(0)} &=&  b_0 \varepsilon_{abc} 
\varepsilon_{def} \ovl{\Psi}_{ad}
          (\Sigma_{be} +i \gamma_5 \Pi_{be}) \Psi_{cf} \no .
\eea
After insertion of the vacuum matrix $\langle M\rangle$, one obtains the 
baryon masses generated by the VEV of the 
two meson fields.  With this kind of coupling, it is 
not possible to describe the correct baryon mass 
splitting as the nucleon and the $\Xi$ are degenerate 
(see table \ref{bartab1}, first column). 
To eliminate this flaw, one can either use 
chirally invariant interaction terms of higher order in the meson fields 
or break the symmetry 
explicitly.\\
Taking the first possibility, one has to compute how the nonlinear 
terms contribute to the baryon masses. 
The {\it quadratic} baryon-meson interaction term 
reads again for the (3,$3^{\ast}$) and ($3^{\ast}$,3) representation 
of the baryons\footnote{Except for the linear term of equation (\ref{bm2}), 
the quadratic and the cubic interactions 
are also possible in the (8,1) and (1,8) representation of the baryons.
Specifically the quadratic contribution reads $\Tr (\ovl{L}MR M^{\dagger}
+\ovl{R}M^{\dagger} LM$). However, this difference will not play a role for 
the vacuum masses or in the 
mean field approximation, since $\langle M \rangle 
=\langle M^{\dagger} \rangle=
\mbox{diag} (\frac{\sigma}{\sqrt{2}} \,, \frac{\sigma}{\sqrt{2}} \, 
,\zeta$).} 
\bea
\label{bm3}
{\cal L}_{BM}^{(1)} &=& b_1\left((\ovl{\Psi}_L)^{\ovl{a}}_b 
M^{b}_{\ovl{c}} (\Psi_R)^{\ovl{c}}_d
             M^d_{\ovl{a}}
          + (\ovl{\Psi}_R)^a_{\ovl{b}} M^{\ovl{b}}_c (\Psi_L)^c_{\ovl{d}}
             M^{\ovl{d}}_a  \right)    \\ \no
         &=& b_1\Tr (\ovl{\Psi}_LM\Psi_RM+\ovl{\Psi}_RM^{\dagger}\Psi_LM^{\dagger}).
\eea
But, as can be observed from table \ref{bartab1}, (second column), 
this term also fails to remove the nucleon- $\Xi$-mass degeneracy. Only the 
inclusion of a {\it cubic} interaction term of the form 
\bea
\label{bm5}
{\cal L}_{BM}^{(2)} &=& b_2\left((\ovl{\Psi}_L)^{\ovl{a}}_b M^{b}_{\ovl{c}} 
(\Psi_R)^{\ovl{c}}_d
             T^d_{\ovl{a}}
          +  (\ovl{\Psi}_R)^{a}_{\ovl{b}} M^{\ovl{b}}_c (\Psi_L)^c_{\ovl{d}}
             T^{\ovl{d}}_a \right)            \\   \no
          &=& b_2\Tr(\ovl{\Psi}_LM\Psi_RT+\ovl{\Psi}_RM^{\dagger}\Psi_LT^{\dagger})\, . 
\eea
yields a mass splitting between nucleon and $\Xi$ (table \ref{bartab1}, 
third column). Here, the dual tensor is defined as\footnote{The 
cubic interaction allows for two independent invariants, the other one being 
analogous to (\ref{bm5})  except for exchanging $T$ and $M$:
\begin{eqnarray*}
{\cal L}_{BM}^{(3)} &=& b_3\left((\ovl{\Psi}_L)^{\ovl{a}}_b T^{b}_{\ovl{c}} 
(\Psi_R)^{\ovl{c}}_d
             M^d_{\ovl{a}}
          +  (\ovl{\Psi}_R)^{a}_{\ovl{b}} T^{\ovl{b}}_c (\Psi_L)^c_{\ovl{d}}
             M^{\ovl{d}}_a\right)            \\   \no
          &=& b_3\Tr(\ovl{\Psi}_LT\Psi_R M+\ovl{\Psi}_RT^{\dagger}\Psi_LM^{\dagger})\, .           
\end{eqnarray*}
However, this form will not be considered, 
because it gives poor results for the baryon mass splitting 
and it does not lead to acceptable nuclear matter fits.} 
\be
T^{\ovl{d}}_a = \epsilon_{a m n} \epsilon^{\ovl{d} \ovl{f} \ovl{g}}M^m_{\ovl 
f} 
M^n_{\ovl g} , 
\ee
so that it transforms in the same way as the meson matrix $M$.\\ 
The second alternative is to 
break the symmetry explicitly. However, the transformation 
properties of the breaking term is restricted due to the 
necessity to maintain the PCAC relation for the pion (section \ref{pcac}).  
Assuming that the mass differences are entirely due to the quark mass 
differences, we break the symmetry along the hypercharge Y-direction. This leads 
to Gell-Mann-Okubo (GMO) mass formulae. We take a term of the 
type (8,8) \cite{sche69},
\be 
\label{sche-esb}
  {\cal L}_{\Delta m} = m_1 \Tr (\ovl{\Psi} \Psi - \ovl{\Psi} \Psi S)
  +m_2 \Tr (\ovl{\Psi} S \Psi)
\ee
where $S_{\ovl{b}}^a = -\frac{1}{3}[\sqrt{3} (\lambda_8)_{\ovl{b}}^a
-\delta_{\ovl{b}}^a]$ (other types as the (8,1) and (1,8), (3,$3^{\ast}$) and 
($3^{\ast}$, 3) representations lead to similar results and are discussed in 
\cite{sche69}). \\
Since none of the baryon-meson interaction terms alone gives the 
correct baryon mass splitting, they will be investigated in combination 
with the explicit symmetry breaking term (\ref{sche-esb}). 
The baryon masses read:
\bea
\label{bmassen}
m_N &=& \frac{b_j}{\sqrt{2}}         B_{j N}  \,\,\, \qquad \quad\qquad \qquad   
m_{\Xi} = \frac{b_j}{\sqrt{2}}     B_{j \Xi}     +m_1  +m_2 \qquad j = 0,1,2  \\ \no
m_{\Lambda} &=& \frac{b_j}{\sqrt{2}} B_{j \Lambda} +\frac{m_1+2 m_2}{3} \qquad
m_{\Sigma} =  \frac{b_j}{\sqrt{2}} B_{j \Sigma}  +m_1 , 
\eea   
with the baryon-meson interaction terms $B_{j \, i}$ ($i=N$, $\Lambda$, $\Sigma$, $\Xi$) of table \ref{bartab1}.
From that one can see that only the strangeness carrying baryon masses are 
modified (note, that in the case of $m_2=m_1\equiv m_{\mbox{s}}$ the 
explicit symmetry breaking term corresponds to  the strange quark mass in the spirit 
of the additive quark model). As only the nucleon 
mass enters the fit to nuclear matter properties, the parameters 
$b_{j}$ shall be fixed to reproduce the nucleon mass. The symmetry 
breaking contributions are then adjusted to the remaining baryon masses. \\
The interaction terms of baryons with spin-0-mesons , which lead to 
a saturating nuclear matter equation of state (see table 
\ref{parameter}), are
\begin{enumerate}
\item L:   ${\cal L}_{BM} = {\cal L}_{BM}^{(0)}+{\cal L}_{\Delta m}$ \quad
      Q:   ${\cal L}_{BM} = {\cal L}_{BM}^{(1)}+{\cal L}_{\Delta m}$ \quad
      C:   ${\cal L}_{BM} = {\cal L}_{BM}^{(2)}+{\cal L}_{\Delta m}$ .
\end{enumerate}
Here, L, Q and C stand for the meson fields entering in the baryon-meson 
interaction terms purely linearly, quadratically and cubic, 
respectively\footnote{The sum of linear, quadratic 
and cubic forms leads also to a realistic baryon mass splitting and 
to a saturating equation of state, even without an explicit symmetry 
breaking term. However, it only 
complicates the discussion without significantly improving results.  
Therefore, we will not consider this option further.} 
(This notation is also used in table \ref{parameter} and in figures 
\ref{energie2}, \ref{felder}, and \ref{baryon}). \\  
 The interaction of the vector meson and axial vector meson nonets 
\bea
V_{\mu}=\frac{1}{\sqrt{2}}\sum_{i=0}^8 v_{\mu}^i\lambda_i \qquad  
A_{\mu}=\frac{1}{\sqrt{2}}\sum_{i=0}^8 a_{\mu}^i\lambda_i
\eea
with baryons is far less involved. 
For the baryons belonging to the $(3,3^{\ast})$ and
($3^{\ast}$,3) representation 
one has the antisymmetric, f-type coupling to baryons\footnote{If the baryons 
are assigned to (8,1) and (1,8),  the analogous octet term reads $g_8^V \Tr( \ovl{\Psi} 
\gamma^{\mu} [V_{\mu},\Psi]
+\ovl{\Psi} \gamma^{\mu} [A_{\mu}\gamma_5,\Psi])$. Since both 
representations differ 
only as to how the axial mesons contribute, there will be no  
difference in the mean field approximation.}  
\be
      {\cal L}_{B V} = g_{8}^V  \Tr( \ovl{\Psi} \gamma^{\mu}[V_{\mu},\Psi]
  +\ovl{\Psi} \gamma^{\mu}\{A_{\mu}\gamma_5,\Psi\})
  + g_1^V \Tr(\ovl{\Psi} \Psi) \gamma^{\mu}\Tr (V_{\mu}+A_{\mu}).
\ee
In the mean field treatment, the axial mesons have a zero VEV. 
 The relevant fields in the SU(2) invariant vacuum, $v^0_{\mu}$ and 
$v^8_{\mu}$, are taken to have the ideal mixing angle 
$\sin \theta_v =\frac{1}{\sqrt{3}}$, yielding
\bea
\label{holz3.36}
  \phi_{\mu}   &=& v_{\mu}^8 \cos \theta_v - v_{\mu}^0 \sin \theta_v =
  \frac{1}{\sqrt{3}} (\sqrt{2} v^0_{\mu}+v_8^{\mu})\\ \no
  \omega_{\mu} &=& v_{\mu}^8 \sin \theta_v + v_{\mu}^0 \cos \theta_v 
 =\frac{1}{\sqrt{3}} (v^0_{\mu}- \sqrt{2} v_8^{\mu}) .  
\eea
For $g_1^V=g_8^V$, the strange vector field $\phi_{\mu} \sim  
\ovl{s}\gamma_{\mu} s $ 
does not couple to the nucleon. The remaining 
couplings to the strange baryons are then determined by symmetry relations:
\be
\label{quarkcoupling}
 g_{\Lambda \omega} = g_{\Sigma \omega} = 2 g_{\Xi \omega} = \frac{2}{3} 
 g_{N \omega}=2 g_8^V \qquad 
 g_{\Lambda \phi} = g_{\Sigma  \phi} = \frac{g_{\Xi \phi}}{2} = 
   \frac{\sqrt{2}}{3} g_{N \omega}  , 
\ee                              
where their relative values are related to the additive quark model. 
In contrast to the baryon/spin-0-meson interaction, two independent 
interaction 
terms of baryons with spin-1 mesons can be constructed. They correspond to 
the antisymmetric (f-type) and symmetric (d-type) couplings, 
respectively. However, from the universality 
principle \cite{saku69} and the vector meson dominance model the d-type coupling should be 
small. 
In mean-field models, large attractive and repulsive contributions from 
scalar and vector mesons cancel to give the relatively shallow nucleon 
potential.
When extended to the strange sector, a different treatment of the 
coupling constants disturbs the cancellation and unphysically 
large hyperon potentials can emerge. We will elaborate on this problem 
in section \ref{hyperon}. 

\subsubsection{Chirally invariant potential}
\label{chiralpot}
The chirally invariant potential includes the mass terms for mesons, their 
self-interaction and the dilaton potential for the breaking of scale 
symmetry. 
For the spin-0 mesonic potential we take all independent combinations of 
mesonic self-interaction terms up to fourth order 
\bea 
\label{mm-pot}
{\cal L}_0\equiv-{\cal V}_0 &= &  \frac{ 1 }{ 2 } k_0 \chi^2 \Tr M^{\dagger} M 
     - k_1 (\Tr M^{\dagger} M)^2 - k_2 \Tr (M^{\dagger} M)^2 \\ \no
     &-&  k_3 \chi ( \det M + \det M^ \dagger ) 
    + k_4 \chi^4 + \frac{1}{4} \chi^4 \ln \frac{ \chi^4 }{ \chi_0^4 } 
-\frac{\delta}{3} \chi^4 \ln \frac{\det M + \det M^{\dagger}}
{2 \det \langle M \rangle} .
\eea
Most of the constants are fixed 
by the vacuum masses of the pseudoscalar and scalar mesons, respectively
(see section \ref{fitsto} for details). 
These are determined by calculating the second derivative of the potential 
in the ground state. Because of the determinant  and the logarithmic terms, mixing 
between $\eta_8$, $\eta_0$  (in the pseudoscalar sector) and $\sigma$, 
$\zeta$, and
$\chi$ (in the scalar sector) occurs, which makes a diagonalization of the 
corresponding mass matrices necessary.   \\
The quadratic and cubic form of the  interaction is made scale invariant 
by multiplying it with an appropriate power of the dilaton 
field $\chi$\cite{cole85}. 
Originally, the dilaton field was introduced by Schechter in order
to mimic the trace anomaly of QCD
$\theta_{\mu}^{\mu}= \frac{ \beta_{QCD} }{2 g} {\cal G}_{\mu \nu}^a {\cal 
G}^{\mu \nu}_a$
in an effective Lagrangian at tree level \cite{sche80} (${\cal G}_{\mu \nu}$
is the gluon field strength tensor of QCD).
The effect\footnote{According to \cite{sche80}, the 
argument 
of the logarithm has to be chirally and parity invariant. This is fulfilled by 
the dilaton which is 
a chiral singlet and a scalar.} of the logarithmic term $ \sim \chi^4 \ln \chi$ is twofold: 
First, it breaks the scale invariance and 
leads to the proportionality $\theta_{\mu}^{\mu} \sim \chi^4$ 
as can be seen from  
\be
\theta_{\mu}^{\mu} = 4 {\cal{L}}  -\chi \frac{\partial {\cal L}}{\partial 
\chi}
- 2 \partial_{\mu} \chi \frac{\partial {\cal L}}
{\partial(\partial_{\mu} \chi)} = \chi^4  , 
\ee
which is a consequence of the definition of scale transformations 
\cite{sche71}. Second, the logarithm leads to a nonvanishing 
vacuum expectation value for the dilaton field resulting in a spontaneous 
chiral symmetry breaking. This connection comes from the multiplication 
of $k_0$ in Eq.~(\ref{mm-pot}) with  $\chi^2$: With the breakdown 
of scale invariance the resulting mass coefficient 
becomes negative for positive $k_0$ and therefore the Nambu--Goldstone 
mode is entered. 
The comparison of the trace anomaly of 
QCD with that of the effective theory allows 
for the identification of the $\chi$-field with the gluon condensate: 
\be
\theta_{\mu}^{\mu} =  \left\langle \frac{ \beta_{QCD} }{2 g} {\cal G}_{\mu \nu}^a 
{\cal G}^{\mu \nu}_a \right\rangle
 \equiv  (1-\delta)\chi^4 .
\ee
The parameter $\delta$ originates from the second logarithmic term with the 
chiral and parity invariant combination $\det M+\det M^{\dagger}$. 
The term is a SU(3)-extension of the logarithmic term proportional to 
$\chi^4 \ln (\sigma^2+ \pi^2)$ introduced in \cite{heid94}. An orientation 
for the value of $\delta$ may be taken from $\beta_{QCD}$ at one loop level, 
with $N_c$ colors and $N_f$ flavors, 
\be
\label{qcdbeta}
   \beta_{QCD}=-\frac{11 N_c g^3}{48 \pi^2} \left(1-\frac{2N_f}{11 N_c}\right)
   +{\cal O}(g^5) , 
\ee                               
where the first number in parentheses arises from the (antiscreening) 
self-interaction of the gluons and the second, proportional to $N_f$, 
is the (screening) contribution of quark pairs. Eq. (\ref{qcdbeta}) suggests 
the value $\delta=6/33$ for three flavors and three colors. This value 
gives the order of magnitude about which the parameter $\delta$ will be varied.\\
For the spin-1 mesons a mass term is needed. The simplest scale invariant 
form
\be
\label{vecfree}
{\cal L}_{vec}^{(1)}= \frac{1}{2} m_V^2 \frac{\chi^2}{\chi_0^2} \Tr (V_{\mu} V^{\mu}
+ A_{\mu} A^{\mu}) 
\ee
implies a mass degeneracy for the meson nonet. To split the masses one can 
add the chiral invariant \cite{gasi69,mitt68}
\be
\label{lvecren}
{\cal L}_{vec}^{(2)} = \frac{1}{8} \mu \Tr[(F_{\mu \nu} +G_{\mu \nu})^2 
M^{\dagger} M +(F_{\mu \nu}- G_{\mu \nu})^2 M^{\dagger} M],
\ee  
with the vectorial and axial field strength tensors $F_{\mu \nu} = 
\partial_{\nu} V_{\mu} -        \partial_{\mu} V_{\nu}$ 
and $G_{\mu \nu} = \partial_{\nu} A_{\mu} -
\partial_{\mu} A_{\nu}$. 
In combination with the kinetic energy term (see Eq. 
\ref{kinetic}),
one obtains for the vector mesons
\bea
\label{kinren}
 &-&\frac{1}{4} [1-\mu \frac{\sigma^2}{2}] (F_{\rho}^{\mu 
\nu})^2
 -\frac{1}{4} [1-\frac{1}{2} \mu (\frac{\sigma^2}{2}+\zeta^2)] 
   (F_{K^{\ast}}^{\mu \nu})^2 \\ \no
 &-&\frac{1}{4} [1-\mu \frac{\sigma^2}{2}](F_{\omega}^{\mu \nu})^2
 -\frac{1}{4} [1- \mu \zeta^2 ] (F_{\phi}^{\mu \nu})^2 .
\eea 
Since the coefficients are no longer unity, the vector meson fields have 
to be renormalized, i.e., the new $\omega$-field reads 
$\omega_r = Z_{\omega}^{-1/2} \omega$.
The renormalization constants are the coefficients  in the square 
brackets in front of the kinetic energy terms of Eq. (\ref{kinren}), 
i.e., 
$Z_{\omega}^{-1} = 1-\mu \sigma^2/2$. The mass terms of the vector mesons 
deviate from the mean mass $m_V $ by the renormalization 
factor\footnote{One could split the $\rho-\omega$ mass degeneracy by adding 
a term of the form \cite{gasi69} 
$ (\Tr F_{\mu \nu})^2$ to Eq. (\ref{kinren}). Alternatively, one could 
break the SU(2) symmetry of the vacuum allowing for a nonvanishing vacuum 
expectation value of the scalar isovector field. 
However,  the $\rho-\omega$ mass splitting is small 
($\sim$ 2 \%), 
we will not consider this complication.}, i.e., 
\be
m_{\omega}^2 = m_{\rho}^2=Z_{\omega} m_V^2 \quad ; \quad   
m_{K^{\ast}}^2 = Z_{K^{\ast}} m_V^2 \quad ; \quad 
m_{\phi}^2 = Z_{\phi} m_V^2 .
\ee
The constants $m_V$ and $\mu$ are fixed to give the correct $\omega$-and 
$\phi$-masses. The other 
vector meson masses are displayed in table \ref{parameter}. 
The axial vector mesons have a mass around 1 GeV. We refrain 
from giving their masses explicitly. To treat them appropriately,  
additional terms are needed \cite{gasi69,ko94}. 
This goes beyond the scope of the present paper. 
\\
\subsubsection{Explicit breaking of chiral symmetry}
\label{pcac}
The term 
\be
{\cal L}_{SB}\equiv-{\cal V}_{SB} = \frac{\chi^2}{\chi_0^2} \Tr (f \Sigma)
= \frac{\chi^2}{\chi_0^2} \left(m_{\pi}^2 f_{\pi}\sigma +
(\sqrt{2}m_K^2 f_K - \frac{ 1 }{ \sqrt{2} } m_{\pi}^2 f_{\pi}) \zeta \right)  
\ee
breaks the chiral symmetry explicitly and makes the pseudoscalar mesons massive\footnote{One 
may wonder why ---besides the explicit symmetry breaking term (\ref{sche-esb}) 
in the baryon-meson sector--- a second chiral noninvariant contribution is 
needed. This is due to our ignorance as to how to transform the current 
quark picture into the constituent quark picture.}. 
It is scaled appropriately to have scale dimension equal to that of the 
quark mass term 
$\sim m_q \overline{q} q+m_s \overline{s}s$, which is present in the 
QCD Lagrangian with massive quarks. 
This term leads to a nonvanishing divergence of the axial 
currents. The matrix elements of $f=1/\sqrt{2}(f_0 \lambda_0+f_8\lambda_8)$ were written as a 
function of $m_{\pi}^2 f_{\pi}$ and $m_K^2 f_K$ to satisfy 
the (approximately valid) PCAC relations for the $\pi$- and $K$-mesons,   
\be
\pa_{\mu} A^{\mu}_{\pi}=m_{\pi}^2 f_{\pi} \pi \, , \qquad 
\pa_{\mu} A^{\mu}_{K} =m_{K}^2 f_{K} K \quad . \no
\ee
Then, by  utilizing the equations of motion,  
the VEV of $\sigma$ and $\zeta$ are fixed in terms of $f_{\pi}$ and $f_K$, i.e.:
\be
   \sigma_0 =-f_{\pi} \qquad \qquad \zeta_0 = \frac{1}{\sqrt{2}}(f_{\pi}-2 f_K) .
\ee 
Since no relation for a partially conserved dilatational current is known, the 
VEV for 
the gluon condensate remains undetermined.    
\subsection{Total Lagrangian}
The kinetic energy terms for the fermions and  mesons are: 
\be
\label{kinetic}
{\cal L}_{kin} = i \Tr \overline{\Psi} \gamma_{\mu} \partial^{\mu}\Psi 
              + \frac{1}{2}\Tr(\partial_{\mu} M^{\dagger} \partial^{\mu} M)
                +\frac {1}{2} \partial_{\mu} \chi \partial^{\mu} \chi 
  - \frac{ 1 }{ 4 } \Tr(F_{ \mu \nu } F^{\mu \nu })  
                - \frac{ 1 }{ 4 } \Tr(G_{ \mu \nu } G^{\mu \nu }) .
\ee
The total general Lagrangian is the sum: 
\be
\label{lagrange}
{\cal L} = {\cal L}_{kin}+{\cal L}_{BM}
+{\cal L}_{BV}+{\cal L}_{vec}+{\cal L}_{0}+{\cal L}_{SB} \no ,
\ee
with ${\cal L}_{vec}={\cal L}_{vec}^{(1)}+{\cal L}_{vec}^{(2)}$. For ${\cal L}_{BM}$, we will discuss the effect of various possibilities 
mentioned in section \ref{bmint} regarding the nuclear matter fits and the 
hyperon potentials.  
%
\subsection{Mean field Lagrangian}
To investigate hadronic matter properties at finite baryon density  we 
adopt the mean-field approximation (see, e.g., \cite{sero86}). 
In this approximation scheme, the fluctuations around constant vacuum 
expectation values of the 
field operators are neglected:
\bea
            \sigma(x)&=&\langle \sigma \rangle +\delta \sigma  
\rightarrow \langle \sigma \rangle \equiv \sigma \, ; \quad 
            \zeta(x)=\langle \zeta \rangle +\delta \zeta  
\rightarrow \langle \zeta \rangle \equiv \zeta \\ \no
      \omega_{\mu}(x) &=&\langle \omega \rangle \delta_{0 \mu}+ 
 \delta \omega_{\mu} 
\rightarrow  \langle \omega_0 \rangle \equiv \omega\, ; \quad 
      \phi_{\mu}(x) =\langle \phi \rangle \delta_{0 \mu}+ 
 \delta \phi_{\mu} 
\rightarrow  \langle \phi_0 \rangle \equiv \phi .
\eea
The fermions are treated as quantum mechanical one-particle operators. 
The derivative terms can be neglected and only the 
time-like component of the vector mesons 
$\omega \equiv \langle \omega_0 \rangle$ and 
$\phi \equiv \langle \phi_0 \rangle$ 
survive if we assume homogeneous and isotropic infinite baryonic
 matter. Additionally, due to 
 parity conservation we have $\langle \pi_i \rangle=0$.
After performing these approximations, the Lagrangian (\ref{lagrange}) 
becomes
\begin{eqnarray*}
{\cal L}_{BM}+{\cal L}_{BV} &=& -\sum_{i} \overline{\psi_{i}}[g_{i 
\omega}\gamma_0 \omega^0 
+g_{i \phi}\gamma_0 \phi^0 +m_i^{\ast} ]\psi_{i} \\ \no
{\cal L}_{vec} &=& \frac{ 1 }{ 2 } m_{\omega}^{2}\frac{\chi^2}{\chi_0^2}\omega^2  
 + \frac{ 1 }{ 2 }  m_{\phi}^{2}\frac{\chi^2}{\chi_0^2} \phi^2\\
{\cal V}_0 &=& \frac{ 1 }{ 2 } k_0 \chi^2 (\sigma^2+\zeta^2) 
- k_1 (\sigma^2+\zeta^2)^2 
     - k_2 ( \frac{ \sigma^4}{ 2 } + \zeta^4) 
     - k_3 \chi \sigma^2 \zeta \\ 
&+& k_4 \chi^4 + \frac{1}{4}\chi^4 \ln \frac{ \chi^4 }{ \chi_0^4}
 -\frac{\delta}{3}\ln \frac{\sigma^2\zeta}{\sigma_0^2 \zeta_0} \\ \no
{\cal V}_{SB} &=& \left(\frac{\chi}{\chi_0}\right)^{2}\left[m_{\pi}^2 f_{\pi} \sigma 
+ (\sqrt{2}m_K^2 f_K - \frac{ 1 }{ \sqrt{2} } m_{\pi}^2 f_{\pi})\zeta 
\right] , 
%
\end{eqnarray*}
with 
the effective mass of the baryon $i$, which 
is defined according to section \ref{bmint}.
\subsubsection{Grand canonical ensemble}
 It is straightforward to write down the expression 
for the thermodynamical potential of the grand canonical 
ensemble $\Omega$ per volume $V$ 
at a given chemical potential $\mu$ and zero temperature:
\be
   \frac{\Omega}{V}= -{\cal L}_{vec} - {\cal L}_0 - {\cal L}_{SB}
-{\cal V}_{vac}- \sum_i \frac{\gamma_i }{(2 \pi)^3}  
\int d^3k [E^{\ast}_i(k)-\mu^{\ast}_i]   
\ee 
The vacuum energy ${\cal V}_{vac}$ (the potential at $\rho=0$) 
has been subtracted in 
order to get a vanishing vacuum energy. $\gamma_i$ denote the fermionic 
spin-isospin degeneracy factors ($\gamma_N=4$, $\gamma_{\Sigma}=6$, 
$\gamma_{\Lambda}=2$, $\gamma_{\Xi}=4$).
The single particle energies are 
$E^{\ast}_i (k) = \sqrt{ k_i^2+{m_i^*}^2}$ 
and the effective chemical potentials read
 $\mu^{\ast}_i = \mu_i-g_{\omega i} \omega-g_{\phi i} \phi$. 
\subsubsection{Equations of motion}
The mesonic fields are determined by extremizing $\frac{\Omega}{V}(\mu, T=0)$:
\bea
\label{tbgls}
\frac{\partial (\Omega/V)}{\partial \chi} &=& 
        -\omega^2 m_{\omega}^2 \frac{\chi}{\chi_0^2} 
        + k_0 \chi (\sigma^2+\zeta^2) 
        - k_3 \sigma^2 \zeta 
        + \left( 4 k_4 + 1 + 4 \ln \frac{ \chi }{\chi_0}
        - 4 \frac{\delta}{3} \ln \frac{\sigma^2 \zeta}{\sigma_0^2\zeta_0}
        \right) \chi^3  \\ 
\no
&+&2\frac{\chi}{\chi_0^2}\left[m_{\pi}^2 f_{\pi}\sigma +(\sqrt{2}m_K^2 f_K - \frac{ 1 }{ \sqrt{2} }
 m_{\pi}^2 f_{\pi}) \zeta \right] =0
\\    \nonumber
\frac{\partial (\Omega/V)}{\partial \sigma} &=& 
 k_0 \chi^2 \sigma - 4 k_1 (\sigma^2+\zeta^2)\sigma 
 - 2k_2 \sigma^3        - 2 k_3 \chi \sigma \zeta  
  -2\frac{\delta \chi^4}{\sigma} \\
&+& \left(\frac{\chi}{\chi_0}\right)^{2} m_{\pi}^2 f_{\pi}  
+ \sum_{i} \frac{\pa m_i^{\ast}}{\pa \sigma}\rho^s_{i}=0 \\  \no      
\frac{\partial (\Omega/V)}{\partial \zeta} &=&  
   k_0 \chi^2 \zeta - 4 k_1 (\sigma^2+\zeta^2) \zeta 
- 4 k_2 \zeta ^3 - k_3 \chi \sigma^2
  -\frac{\delta \chi^4}{\zeta} \\
 &+& \left(\frac{\chi}{\chi_0}\right)^{2}  \left[\sqrt{2}m_K^2 f_K 
- \frac{ 1 }{ \sqrt{2}} m_{\pi}^2 f_{\pi}\right]+\sum_{i} 
\frac{\pa m_i^{\ast}}{\pa \zeta} \rho_i^s=0 .
\eea
The vector fields $\omega$ and $\phi$ are determined from  
$\frac{\partial (\Omega/V)}{\partial \omega}=0$  and  
$\frac{\partial (\Omega/V)}{\partial \phi}=0$, respectively. They may be 
solved
explicitly yielding
\begin{equation}
\omega=\frac{g_{i \omega } \rho^i \chi_0^2}{m_{\omega}^2 \chi^2}, \qquad  
\phi=\frac{g_{i \phi } \rho^i \chi_0^2}{m_{\phi}^2 \chi^2} .
\end{equation}
The scalar densities  $\rho^s_{i}$ and the vector densities $\rho_i$ can 
be calculated analytically, yielding
\bea
\rho^s_i &=& \gamma_i
\int \frac{d^3 k}{(2 \pi)^3} \frac{m_i^{\ast}}{E^{\ast}_i} = 
\frac{\gamma_i  m_i^{\ast}}{4 \pi^2}\left[ k_{F i} E_{F i}^{\ast}-m_i^{\ast 2} 
\ln\left(\frac{k_{F i}+E_{F i}^{\ast}}{m_i^{\ast}}\right)\right] \\ \no
 \rho_i &=&  \gamma_i \int_0^{k_{F i}} \frac{d^3 k}{(2 \pi)^3} =
\frac{\gamma_i k_{F i}^3}{6 \pi^2}        \,   .
\eea
The energy density and the pressure  follow from the Gibbs--Duhem relation, 
$\epsilon = \Omega/V+ \mu_i \rho^i$ and $p= - \Omega/V$. 
Applying the Hugenholtz--van Hove theorem \cite{hugo58}, the Fermi surfaces 
are given by $ E^{\ast}(k_{F i})= \sqrt{k_{F i}^2+m_i^{\ast 2}} 
= \mu^{\ast}_i $ . 
           
\section{Results}
The scope of the present paper is to explore whether it is possible 
to describe nuclear-matter properties reasonably well within the framework 
of the $SU(3)_L \times SU(3)_R$ $\sigma$-model. Therefore, we discuss 
only the results for the limit of vanishing {\it net} strangeness. 
The case of finite strangeness will be discussed in a forthcoming 
publication \cite{paper4}. 
However, there are strong implications of the Lagrangian for the hyperon 
potentials and for high densities, which will be elaborated in the 
following.    
\subsection{Fits to nuclear matter and the hadron masses}
\label{fitsto}

A salient feature of all chiral models are the strong vacuum constraints. 
In the present case they fix  $k_0$, $k_2$ and $k_4$,  in
order to minimize the thermodynamical potential $\Omega$ in vacuum
for given values of the fields $\sigma_0$, $\zeta_0$ and $\chi_0$. Note that
these parameters could also be eliminated by adding appropriate chirally 
invariant terms
to ensure that the vacuum energy is minimal for given values of $\sigma_0$, 
$\zeta_0$ and $\chi_0$.
The parameter $k_3$ is  fixed to the $\eta$-mass m$_{\eta}$. 
There is some freedom to vary parameters, mainly due to 
the unknown mass of the $\sigma$-meson, $m_{\sigma}$, 
which is determined by $k_1$, 
and due to the uncertainty of the value for the 
kaon decay constant $f_{K}$. 
While the kaon decay constant is not known precisely, the value for 
f$_{\pi}$ is known very well. Hence, we keep f$_{\pi}$ fixed to 93 MeV
and vary $f_K$ in the range 115$\pm 5$~MeV.\\
In order to reproduce the correct nuclear matter properties, two 
of the parameters have to be adjusted to the medium.
We choose $g_{N \omega}$ and 
$\chi_0$ to fit
the binding energy of nuclear matter $\epsilon/\rho_B-m_N=-16$~MeV at the saturation 
density 
$\rho_0 =0.15 \, \rm fm^{-3}$.
It should be noted that a reasonable nuclear matter fit with acceptable 
compressibility \cite{kuon95} can be found 
(row L in table \ref{parameter}), where $m_{\sigma} \approx$ 500 MeV. 
This, in the present approach, allows for an interpretation of the $\sigma$-field
as the chiral partner of the $\pi$-field and as the mediator  
of the mid-range attractive force between nucleons, though we believe 
the phenomenon is in reality  
generated through correlated two-pion exchange \cite{sero97}.\\
Generally, the fits of table 
\ref{parameter} have an effective nucleon mass of $m_N^{\ast}=(0.7-0.75) m_N$ 
and a compressibility of about 300 MeV. Although these values are 
reasonable, it might be desirable to fine tune them in order 
to get acceptable fits to nuclei, too. This could be done 
by adding a quartic self-interaction of spin-1 mesons, e.g., 
$\Tr \left[ (V_{\mu}+A_{\mu})^4+(V_{\mu}-A_{\mu})^4 
\right]$.



\subsection{Hyperon potentials}
\label{hyperon}
Besides the observables pressure $p$, energy per baryon $\epsilon/\rho_B$,
compressibility $K$ and effective nucleon mass $m_N^{\ast}/m_N$ at ground 
state density $\rho_0$, there are some additional important
constraints in the medium due to hypernuclear physics: 
The (relativistic) potential depths $U_i$ of the 
baryons at $\rho_0$, which can serve as input to restrict also the 
`nonstrange' parameters,  
\be
\label{poti}
U_{i} = m_{i}^{\ast}-m_{i}+g_{\omega i}\omega ,  
\quad  i = N, \Lambda, \Sigma, \Xi .
\ee
Experimentally, one finds for 
the $\Lambda$-hyperons a potential of U$_{\Lambda}$ =$-30\pm$3 MeV
 \cite{Dover84}.
 For the $\Sigma$-potential the situation is unclear, 
since there is no evidence for bound $\Sigma$-hypernuclei. The 
predictions range from completely unbound $\Sigma$'s  \cite{ma95} to 
U$_{\Sigma} = -25\pm$5 MeV \cite{Dov89a}. For $\Xi$-hyperons, several bound 
$\Xi$-hypernuclei candidates have been reported \cite{Dover84}. 
The potential for 
the $\Xi$-hyperon has been extracted to $U_{\Xi} = -25\pm5$ MeV. \\
The Yukawa-type chirally invariant baryon-meson interaction
gives an acceptable mass spectrum of mesons and baryons 
(row L of table \ref{parameter}), and also, the compressibility
has a reasonable value (K $\approx$  300 MeV). However, 
the potential depths of the hyperons are very deep.  
This is mainly due to the baryon-vector and baryon-scalar meson 
coupling constants, 
which determine the strength of the vector and scalar potential, respectively
(see Eq. \ref{poti}). Once $g_{N\sigma}$ and $g_{N\omega}$ are fixed to the 
nucleon mass 
and the nuclear potential, the coupling constants of strange baryons 
to mesons are determined by symmetry relations.  
As discussed in section \ref{bmint}, chiral symmetry restricts
the coupling of spin-0 mesons to baryons to a symmetric (d-type) one. 
This destroys the 
balance between repulsion and attraction, since the baryon-vector coupling 
is antisymmetric (f-type), i.e. $g_{\Sigma \sigma}$=0, whereas 
$g_{\Sigma \omega}=\frac{2}{3} g_{N \omega}$.
To cure this deficiency, nonlinear baryon-meson
 interaction terms can be introduced, which are also chirally invariant. 
 They lead to coupling constants which differ from 
the Yukawa-type baryon-meson interaction. \\
 The results of fits to nuclear matter are shown in the 
rows 2-4 of table \ref{parameter}. 
If quadratic baryon-meson interactions (Q-fit) are used, the 
hyperon potentials 
are still too deep. {\it Cubic} baryon meson interactions (C-fit) 
allow for a coupling of the strange condensate to the nucleon, such that 
all baryon potentials are acceptable. This is because the scalar coupling 
constants approach those for the f-type coupling (Eq. \ref{quarkcoupling}). 
This implies that nonstrange mesons couple according to the OZI rule, i.e. 
exclusively to the up and down quark, but not to the strange quark. With 
such a coupling scheme, hypernuclei can be reasonably well
described\cite{jennings90}.                                                                                      
The potentials of the $\Sigma$- and $\Lambda$-hyperons are then equal 
since 
their density-dependent mass terms are the same (see fourth column of 
table \ref{bartab1}). It is remarkable that in this nonlinear scheme 
a coupling of the strange 
condensate to nucleons is necessary to yield a $\Lambda$-potential of 
the right magnitude.\\
Other possibilities than the cubic form 
of baryon-meson interaction may
exist to yield realistic hyperon potentials. 
The explicit symmetry breaking term in Eq. (\ref{sche-esb}) 
has no influence on the potential, since it is not medium dependent. 
Other forms of explicit symmetry breaking, which involve the meson fields (they are listed in \cite{sche69}), either fail to 
generate the 
experimentally known baryon mass spectrum or give unrealistically 
high/low potentials\footnote{If 
one includes 
four instead of two parameters, then it is of course possible to fit both 
the potentials {\it and} the 
baryon masses simultaneously, however for the price of 
losing the predictive power.}. \\
We have also checked the inclusion of a d-type coupling of baryons to spin-1 mesons. 
Then, the couplings in the baryon-scalar and baryon-vector meson 
sector can be chosen to be of the 
same magnitude. Indeed, a pure d-coupling of baryons to spin-1 mesons 
leads to acceptable hyperon potentials!
However, this yields negative couplings of nucleons to the $\rho$-meson, 
in contrast to experiment. 
Correcting this deficiency by adding chiral symmetry breaking 
terms into baryon-vector sector seems artificial  and it does not correct
 the contradiction to vector meson dominance and to the 
universality principle \cite{bern68,saku69}. \\
The nonlinear (cubic) baryon--meson interaction term that gives 
reasonable hyperon potentials (row C in table 
\ref{parameter}) can be considered as an effective  description of baryons 
interacting with multi--quark states. This interpretation is analogous to 
the common view of the $\sigma$-meson in the one-boson-exchange models 
as a effective parameterization of the correlated two--pion exchange. 

\subsection{Equation of state and effective baryon masses}

\begin{figure}
\epsfbox{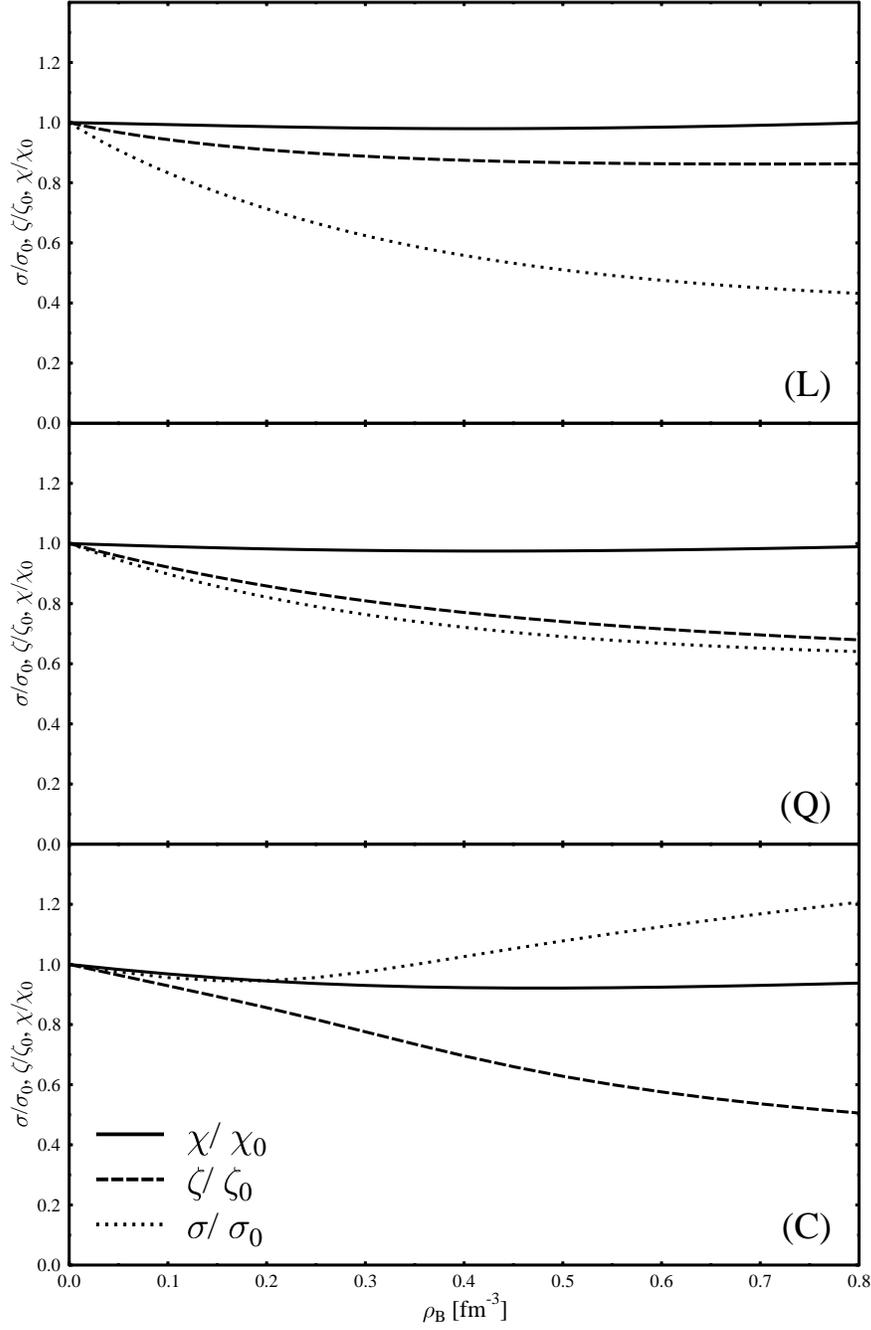}
\caption{\label{felder} Nonstrange ($\sigma$), strange ($\zeta$) and gluon ($\chi$) S
condensates versus baryon density $\rho_B$
for the 
linear (L), quadratic (Q), and cubic (C) baryon-spin-0 meson interaction 
(see table \ref{parameter}). }
\end{figure}

In spite of all the differences in the baryon-meson interaction, 
all fits of table \ref{parameter} lead to almost\footnote{At higher baryon densities 
the fits L, Q, and C deviate from each other, since their compressibilities are 
slightly different.} identical  
equations of state (see Fig. \ref{energie2}).  
In contrast, 
the density dependence of the condensates is characteristic 
for the specific form of the chirally invariant baryon-spin-0 meson 
coupling used (Fig. \ref{felder}).\\
If the baryons are coupled linearly to 
spin-0 mesons, the (nonstrange) $\sigma$- field decreases linearly 
at low densities, 
and then it saturates at nearly 40 \% of its VEV 
(Fig. \ref{felder}L). 
This behavior is in contrast to the linear Walecka model where 
$m_N^{\ast} \rightarrow 0$.   
The strange condensate $\zeta$ changes only slightly 
in the nuclear medium, since it does not couple to the scalar density 
of nucleons.\\
This is different for the quadratic (Fig. \ref{felder}Q) 
and cubic (Fig. \ref{felder}C) forms of baryon-meson interaction. 
There, the strange and nonstrange fields couple either equally to the 
nucleons (see column Q of table \ref{bartab1}), or even stronger than $\sigma$ 
(column C of table \ref{bartab1}). Consequently, the medium dependence 
of the strange condensate becomes stronger, and that of $\sigma$ weakens 
with increasing the nonlinearity of the baryon-scalar meson coupling. 
The dilaton $\chi$ changes negligible in the medium for all kinds of 
interaction terms, since it corresponds to a heavy ($>1$ GeV) particle, 
and it does not couple to the scalar density of nucleons.\\
The baryon masses are generated dynamically through the strange and 
nonstrange condensates. Therefore, they are density dependent, too 
(Fig. \ref{baryon}). Their medium behavior follows from that 
of the condensates and from the chirally invariant `mass terms' 
of table \ref{bartab1}. At high densities, the masses saturate (or even 
increase), in contrast to the masses in the 
Walecka model, which drop dramatically. The main difference between the various fits 
is the density dependence of the mass of the $\Xi$-hyperon, which is weakest for 
the cubic baryon-meson interaction, since there it couples only to $\sigma$.
 
\subsection{Chiral symmetry restoration}  

\begin{figure}
\epsfbox{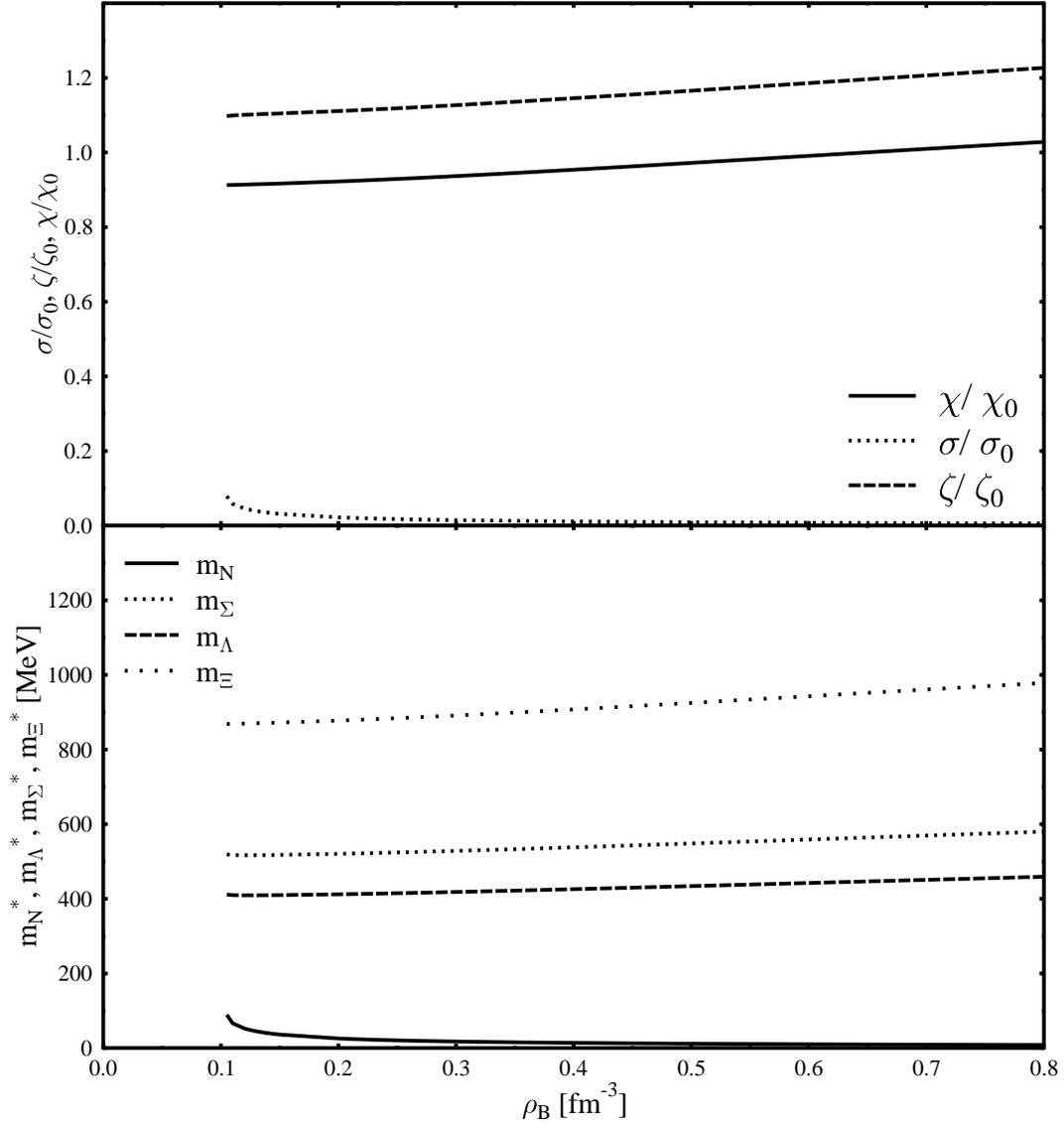}
\caption{\label{abnormal} Nonstrange ($\sigma$), 
strange ($\zeta$) and gluon ($\chi$) condensates (above) and 
effective baryon masses (below) in the abnormal 
(chiral) phase.}
\end{figure}

Although the Lagrangian with the three different types of baryon-meson 
interaction is chirally invariant, there is no 
chiral phase transition at high baryon densities. This is not a deficit 
of our (purely hadronic) model, since at very high densities 
the mean-field model with parameters fixed at $\rho_0$ is most probably out 
of its range of applicability. Furthermore, it is unclear, whether 
a chiral symmetry restoration at high densities takes place or not \cite{enhance}.\\
However, in the chiral $\sigma-\omega$ model, a solution besides the one 
describing normal nuclear matter can be found, which has the features 
of a chirally restored phase with, e.g., a vanishing effective nucleon mass.
This abnormal solution exists only for a certain range of parameters. 
As pointed out in \cite{paper1}, the abnormal phase does only 
exist, if the Lagrangian does not include terms 
which lead to a contribution in the equation of motion proportional 
to $1/\sigma$ or higher powers of it. 
The logarithmic term $\sim \ln \det \sigma^2 \zeta$ is such an example. 
For the linear baryon--meson interaction, the absence of such a term 
leads to an unrealistically large nuclear matter compressibility of 
K $\approx 1400$ MeV \cite{paper1}. This is not the case for the 
cubic baryon-meson interaction. There, even with $\delta=0$, the 
compressibility is about $K\approx  300$ MeV. Therefore, the nonlinear 
coupling of baryons to scalar mesons reduces the compressibility as 
compared to the Yukawa-type of coupling and makes the equations of state softer.
However, the abnormal solution following from such a fit 
is absolutely stable even at $\rho_0$.   
It is possible to shift the abnormal phase to higher energies, 
so that it becomes metastable, if the term 
\be
{\cal L}_{vec}^2 = g_2 \Tr\left[(V_{\mu}+A_{\mu})M M^{\dagger}(V^{\mu}+A^{\mu}) 
+(V_{\mu}-A_{\mu})M^{\dagger} M (V^{\mu}-A^{\mu}) \right]
\ee
is included, and the effective $\omega$-meson mass is generated predominantly 
by $\sigma$, e.g.    
\be
\omega=\frac{g_{i \omega } \rho^i}{m_V^2\chi^2/\chi_0^2+g_2\sigma^2 }
\ee
where $m_V$ and $g_2$ are fixed to the masses $m_{\omega}$ and 
$m_{\phi}$ (here, the renormalization of 
the $\omega$-field is  neglected by setting $\mu=0$, see Eq. \ref{lvecren}). 
A fit with $g_2=30.0$ and $m_V=594.7 $ MeV, a reasonable compressibility and 
realistic hyperon potentials is given in table \ref{parameter} ($C_a$-fit, $a$ 
stands for abnormal\footnote{For a correct description of the axial 
vector meson mass splitting, a term of the 
form $\Tr[(V_{\mu}+A_{\mu}) M (V^{\mu}-A^{\mu}) M^{\dagger}]$ should be added.}).
As shown in Fig. \ref{energie2}, the abnormal phase 
of nearly massless nucleons has ---at zero net strangeness---
always a higher energy than the phase describing normal nuclear matter. 
In contrast to the SU(2)--equation of state \cite{mish93,paper1}, 
the abnormal and normal branch do not cross each other, so that no 
phase transition occurs at high baryon densities. 
Nevertheless, it is instructive to look at the condensates and the baryonic 
masses of the abnormal phase: 
Although the onset of a phase transition is highly parameter dependent, 
the features of the abnormal or chirally restored phase are not.\\   
In contrast to the nearly vanishing $\sigma$ field, 
the strange scalar field $\zeta$ has a high value in the abnormal phase
(Fig. \ref{abnormal}). This is connected to the absence of 
repulsion in the strange sector: There is no contribution from 
the $\omega$ (since it does not couple to the strange condensate), and 
the $\phi $ (which depends on $\zeta$) does not couple 
to the nucleon density.
In the abnormal (chiral) phase not all baryon masses vanish. Their mass difference is 
due to the explicit symmetry breaking term (\ref{sche-esb}). \\
A thorough analysis of the parameter dependence and the onset of the 
chiral phase transition at high densities and nonzero strangeness fraction 
will be postponed until finite nuclei are described satisfactorily 
with the cubic baryon-meson interaction\cite{paper3}.\\
Although the cubic baryon--spin--0 meson interaction term gives 
reasonable results for infinite nuclear matter, it seems 
a rather artificial construction.
The question still remains as to whether it is possible to 
keep both the Yukawa-type
baryon-meson interaction and at the same time to yield reasonable hyperon potentials in a 
chiral model.  
A model which gives a positive answer is proposed in the following 
section.   
\section{A model with hidden chiral symmetry}
The difficulties encountered when chirally invariant baryon-meson 
interactions are introduced is presumably related to the large mass of 
the baryons as compared 
to the mass of the pion. At this energy scale chiral symmetry is known 
to be a useful concept. 
A general framework on how to add `heavy particles' without destroying chiral 
symmetry 
was presented in the classic papers of refs. \cite{wein68,cole69,callen69}. 
The idea is to go over to a representation where the heavy particles
transform equally under left and right rotations. To accomplish 
this, it is 
necessary to dress these particles nonlinearly with pseudoscalar mesons. 
The application of this     
method to our approach has the following advantages:
\begin{itemize}
 \item the Yukawa-type baryon/spin--0 meson interaction can be retained, 
 \item the strange baryons have reasonable potential depths,
 \item the heavy particles transform in the $SU(3)_V$ space, i.e., their 
       interaction terms are not restricted by chiral symmetry, which is 
       expected to hold mostly for light particles,  
\item baryon masses can be fitted without explicit symmetry breaking 
 terms,
 \item a connection to the phenomenologically successful Walecka-model 
  exists.
\end{itemize}
In the following we will outline the argumentation. 
For a thorough discussion, see \cite{paper3}.\\
Let the elementary spinors (=quarks) $q$ introduced in section \ref{repr} 
transform into
'new' quarks $\tilde{q}$ by
\be
\label{nlq}
      q_L(x) = U(x) \tilde q_L(x) \qquad q_R(x) = U^{\dagger}(x) \tilde q_R(x)  
\ee
with the pseudoscalar octet $\pi_a$ arranged in $U(x) = \exp[-i\pi_a 
\lambda^a/2]$. 
Since the algebraic 
composition of mesons in terms of quarks is known (see section \ref{mesons}), 
it is 
straightforward to transform form `old' mesons $\Sigma$ and $\Pi$ 
into `new' mesons $X$ and $Y$:
\be
  M = \Sigma+i\Pi = U (X+iY)U \quad .
\ee
Here, the parity even part $X$ is associated with the scalar nonet, whereas 
$Y$ 
is taken to be the pseudoscalar singlet \cite{stoks96}. 
In a similar way, the `old' baryon octet 
 $\Psi$ forming the representation (8,1) and (1,8) 
is transformed into a `new' baryon octet  $B$:
\be
 \Psi_L = U B_L U^{\dagger} \qquad \Psi_R = U^{\dagger} B_R U \quad .
\ee
The transformations of the exponential $U$ are known 
\cite{cole69,callen69}, 
\be
     U' = LUV^{\dagger} = VUR^{\dagger}, 
\ee
and with the `old' fields from table \ref{chiraltrafo}, 
the `new' baryons $B$ and the `new' scalar mesons $X$  
transform as\footnote{For vector transformations we have L=R=V, whereas for $L \neq R$,  
$V$ 
is a complicated nonlinear function of the pseudoscalars $\pi_a(x)$}:
\begin{eqnarray}
 B'_L &=& VU^{\dagger}  L^{\dagger} \cdot L\Psi_L L^{\dagger} \cdot L U V^{\dagger} = VB_LV^{\dagger}\\ \no
 B'_R &=& VU R^{\dagger} \cdot R \Psi_R R^{\dagger} \cdot R U^{\dagger} V^{\dagger} = VB_RV^{\dagger}\\
 X'   &=& \frac{1}{2}(VU^{\dagger}  L^{\dagger} \cdot L M R^{\dagger} \cdot 
      R U^{\dagger} V^{\dagger} +
            VU R^{\dagger} \cdot R M^{\dagger} L^{\dagger}  \cdot 
                       L U V^{\dagger}) = V X V^{\dagger}     \no
\end{eqnarray}
The pseudoscalars reappear in the transformed model 
as the parameters of the symmetry transformation.  
Therefore,  chiral invariants (without space-time derivatives) are  
independent of the Goldstone bosons. Hence, in mean field approximation, 
the potential (\ref{mm-pot}) does not change its form (see also \cite{bwlee69}).
Furthermore, the `new' fields allow for invariants which are forbidden for 
the `old' fields by chiral symmetry:
Since the baryons and scalar mesons now transform equally in the left and 
right subspace, the f-type coupling for the baryon-meson interaction
is now allowed. The invariant linear interaction terms of baryons to 
scalar mesons are
\be
{\cal L}_{BX} = g_8^S \left(\alpha[\ovl{B}BX]_F+ (1-\alpha)  [\ovl{B} B X]_D 
\right)
+ g_1^S \Tr(\ovl{B} B)\Tr X  \, ,  
\ee
with $[\ovl{B}BX]_F:=\Tr(\ovl{B}BX-\ovl{B}XB)$ and 
$[\ovl{B}BX]_D:= \Tr(\ovl{B}BX+\ovl{B}XB) - \frac{2}{3} 
\Tr (\ovl{B} B) \Tr M$. In contrast to table 
\ref{bartab1} (first column), the baryon masses 
have an additional dependence on $\alpha$:
\bea
\label{bmassen2}
 m_N  &=& m_0 -\frac{1}{3}g_8^S(4\alpha-1)(\sqrt 2\zeta-\sigma) \\ \no
 m_{\Lambda}&=& m_0-\frac{2}{3}g_8^S(\alpha-1)(\sqrt 2\zeta-\sigma) \\ \no
 m_{\Sigma} &=& m_0+\frac{2}{3}g_8^S(\alpha-1)(\sqrt 2\zeta-\sigma)  \\ \no
 m_{\Xi}    &=& m_0+\frac{1}{3}g_8^S(2\alpha+1)(\sqrt 2 \zeta-\sigma) \no
\eea
with $m_0=g_1^S(\sqrt{2} \sigma+\zeta)/\sqrt{3}$.
The three parameters $g_1^S$, $g_8^S$ and $\alpha$ can be used to fit the baryon masses 
to their experimental values. Then, no additional explicit symmetry breaking term 
is needed.
For $\alpha=0$ and $g_1^S=-\sqrt{2/3}g_8^S$, 
the d-type coupling of table \ref{bartab1} is recovered, and 
for $\zeta=\sigma/\sqrt{2}$ (i.e. 
$f_{\pi}=f_K$), the masses are degenerate, and the vacuum is $SU(3)_V$ 
invariant.
The potentials following from the fit to nuclear matter are for 
$\alpha=1.13$: 
$U_N=-58.4$ MeV, $U_{\Lambda}=-39.5$ MeV, $U_{\Sigma}=-30.0$ MeV, and 
$U_{\Xi}=-15.8$ MeV. Note that the sum $U_{\Lambda}+U_{\Sigma}$ is 
independent of the mixing angle $\alpha$ (this can be seen by inserting 
Eq. (\ref{bmassen2}) in Eq. (\ref{poti})). As in the cubic fit, 
a coupling of the strange condensate to the nucleon is necessary to 
obtain acceptable potential depths. \\
Since the construction of invariants is only governed by $SU(3)_V$, 
the form of the `new' Lagrangian is analogous to the one used in 
RMF-models\cite{sero86} and allows for equally good 
results when applied e.g. to finite nuclei \cite{thera2,paper3}. 
In contrast to the Walecka model relations following from chiral symmetry as 
PCAC and the Goldberger-Treiman relation are incorporated. The model allows 
also to predict the masses of the meson nonet at zero and finite 
density\cite{paper4}.   

\section{Summary and Outlook}
We have presented a chiral $SU(3)_L \times SU(3)_R$ linear $\sigma$ model for 
finite baryon density.
Besides the meson-meson interaction, which is widely used 
\cite{levy67,gasi69,sche71,sche80}, 
spin-1 mesons and baryons with dynamically generated masses are implemented. 
In addition, a dilaton field is used to render the Lagrangian scale invariant, 
except for a scale breaking logarithmic term which simulates the 
trace anomaly of QCD.\\ 
The parameters are fixed to the hadron masses and to the binding energy of 
nuclear matter at zero pressure. 
These parameters can all be related to and are constrained by physical quantities.
The equation of state of nuclear matter then has a compressibility constant 
of about 300 MeV. 
Nevertheless, the extension to SU(3) is nontrivial, because of the 
constraints imposed by chiral symmetry on the baryon-meson interaction. 
The linear form of the interaction leads 
to coupling constants given by the $d_{ijk}$-structure constants. 
Combined with 
the baryon-vector interaction, which go like $f_{ijk}$, they generate 
false hyperon potentials. This problem can be circumvented by using 
a cubic baryon-meson interaction, whose coupling constants are similar 
to the $f$-type ones.\\
 Another possible way out of this dilemma (and maybe more natural) 
 is the nonlinear realization of the $\sigma$-model\cite{wein68,cole69}. 
 With a nonlinear transformation into 
`new' scalar fields transforming linearly in $SU(3)_V$ and 
into `new' pseudoscalar fields transforming nonlinearly, it is possible to 
construct an f-type baryon--scalar meson interaction. The mixing angle 
between $d$ and $f$ can then be used to adjust to the known potential 
of the $\Lambda$-hyperon.  
Furthermore, no additional explicit 
symmetry breaking mass term for the baryons is needed. The modified form 
of the Lagrangian can be recast to resemble the nonlinear Boguta--Walecka
Lagrangian of ref. \cite{scha93}, which was successfully applied to finite 
nuclei and hypernuclei. A thorough investigation of this modified model 
and its connection to the nonchiral mean-field models is presently under way\cite{paper3}.\\
It is found that both in the cubic 
form of baryon-scalar meson interaction and in the nonlinear realization of 
chiral symmetry, the strange condensate needs to be coupled 
to the nucleon in order to obtain realistic hyperon potentials. This 
may be viewed as for a large strangeness content of the nucleon\cite{strangenuc}.\\
The cubic model
($C_a$), allows for an abnormal `Lee--Wick' phase with nucleons of 
nearly vanishing mass. In contrast to SU(2) models 
involving an abnormal phase \cite{paper1}, here the normal phase, 
which describes ordinary
nuclear matter, has a reasonable compression modulus 
(K $\approx 300$ MeV). 
In the abnormal phase, the strange condensate remains ---in contrast to 
the (vanishing) $\sigma$ field--- close to its VEV. \\
The case of zero net strangeness as well as
the effective baryon masses at higher densities were studied here. In a forthcoming 
publication \cite{paper3} the extension to finite strangeness and the 
behavior of the meson masses in matter will be discussed in detail. 
The application of the model to finite nuclei is currently 
under investigation.
  
\begin{acknowledgements}
The authors are grateful to J. Eisenberg, C. Greiner,  
K. Sailer and D. Zschiesche for fruitful discussions. This work 
was funded in part by Deutsche Forschungsgemeinschaft (DFG), Gesellschaft 
f\"ur Schwerionenforschung (GSI) and 
Bundesministerium f\"ur Bildung und Forschung (BMBF). J. Schaffner-Bielich
is supported by a 
Feodor Lynen fellowship from the Alexander von Humboldt-Stiftung. 
\end{acknowledgements}


%

%
%
\pagebreak
\begin{table}
\caption{Chiral transformations of  spin-0 mesons 
($M=\Sigma+i\Pi$), spin-1 mesons ($V_{\mu}=l_{\mu}+r_{\mu}$ and 
$A_{\mu}=l_{\mu}-r_{\mu}$) and baryons (see Eq. \ref{barlr}).}
\label{chiraltrafo} 
\begin{tabular}{lccc}  
 Hadrons                      & $J^{P}$               & 
\multicolumn{2}{c}{Transformations}      \\ \hline 
Spin-0 mesons           & $0^+, 0^-$             &  $LM R^{\dagger}   $                 
& $RM^{\dagger} L^{\dagger} $ \\ 
Spin-1 mesons          & $1^-, 1^+$        & $L l_{\mu}  L^{\dagger}$ 
 & $R r_{\mu} R^{\dagger}$ \\ 
baryons (nonet)           & $\frac{1}{2}^+$ &   $L\Psi_L R^{\dagger} $        
& $R\Psi_R L^{\dagger} $ \\ 
baryons (octet)             & $\frac{1}{2}^+$ & $L \Psi_L  L^{\dagger}$  & 
$R \Psi_R R^{\dagger}$ \\ 
\end{tabular}
\end{table}
 
\begin{table}
\caption{\label{bartab1}Mass terms $B_{ji}$ for the baryons 
(see Eq. \ref{bmassen}).}
\begin{tabular}{cccc} 
        & $j=0$          & $j=1$              & $j=2$  \\ \hline
N       & $\sigma$      & $\sigma \zeta$    &$2 \sigma \zeta^2$  \\
$\Xi$     & $\sigma$      &     $\sigma \zeta$  &$\sigma^3$      \\
$\Lambda$ & $\frac{1}{3}(4 \sigma-\sqrt{2} \zeta)$ & 
          $\frac{\sqrt{2}}{6}( \sigma^2+4 \zeta^2)$ 
&$\sqrt{2} \sigma^2 \zeta$ \\
 $\Sigma        $& $\sqrt{2} \zeta$ & $\frac{1}{\sqrt{2}} \sigma^2$ & 
 $\sqrt{2} \sigma^2 \zeta$  \\ 
\end{tabular}
\end{table}
\begin{table}
\caption{\label{parameter} Parameterizations with linear (L), quadratic 
(Q), cubic baryon-meson interaction with ($C_a$) and without (C) abnormal (chiral) 
phase. All masses are given in MeV.}
\bt{ccccccccc}
 & \multicolumn{8}{c}{Spin-0 particle masses} \\ 
 & $m_{\pi}$(139) & $m_{K}$(495) & $m_{\eta}$(547) & $m_{\eta'}$(958) & $m_{\sigma_{\pi}}$  & $m_{\sigma_{K}}$  & $m_{\sigma_{\eta'}}$ & $m_{\sigma_{\eta}}$ \\ \hline 
 L      & 139.0 & 498.0 & 540.0 & 931.1 & 973.3    & 1065.0 & 561.0 & 747.1 \\ 
 Q      & 139.0 & 498.0 & 540.0 & 972.4 & 1064.8   & 1169.3 & 728.9 & 993.1 \\
 C      & 139.0 & 498.0 & 540.0 & 954.6 & 1023.4   & 1122.8 & 774.0 & 1056.6 \\
 C$_a$  & 139.0 & 495.0 & 540.0 & 946.0 & 966.8    & 1041.5 & 665.5 & 968.6 \\ \hline
   &\multicolumn{4}{c}{Spin-1 particle masses}  &\multicolumn{4}{c}{Spin-$\frac{1}{2}$ particle masses} \\ 
  & m$_{\omega}$(783) & m$_{K^{\ast}}$(892) &  m$_{\rho}$(770)& m$_{\phi}(1020)$ & m$_N$(939)  & m$_{\Lambda}$(1115) & $m_{\Sigma}$(1193) & m$_{\Xi}$(1315) \\ \hline
 L     & 783.0  & 878.4  & 783.0 & 1020.0 & 939.0  & 1104.9 & 1193.1 & 1314.9 \\ 
 Q     & 783.0  & 878.4  & 783.0 & 1020.0 & 939.0  & 1177.3 & 1198.5 & 1314.9 \\
 C     & 783.0  & 864.0  & 783.0 & 1020.0 & 939.0  & 1071.5 & 1159.5 & 1339.0 \\
 C$_a$ & 783.0  & 867.5  & 783.0 & 983.4  & 939.0 & 1057.0  & 1174.7 & 1348.7 \\ \hline
 & \multicolumn{4}{c}{Potential depths [MeV]} &  & & & \\ 
& U$_N$   & U$_{\Lambda}$ & U$_{\Sigma}$ & U$_{\Xi}$ & $m_1$ [MeV]& $m_2 $[MeV] &
$\frac{m_N^{\ast}}{m_N}$ & K [MeV] \\ \hline 
 L     & -60.6 & -149.9 &-2.5   & -165.5 & -230.5 & 606.4  & 0.77   & 279.3 \\
 Q     & -60.9 & -136.4  &-68.0 & -169.7 & 531.9  & -156.0 & 0.76   & 313.6 \\ 
 C     & -62.1 & -27.9  & -27.9 & -28.5  & 512.6  & 380.7  & 0.74   & 329.9 \\ 
 C$_a$ & -61.3 & -29.8  &-35.0  & -47.0  & 535.3  & 373.2  & 0.76   & 306.3  \\ \hline
 & \multicolumn{8}{c}{Parameter}\\ 
 & $33 \delta$   & g$_{N \omega}$ & k$_0$ & k$_1$ & k$_2$  & k$_3$  & k$_4$ & f$_K$ [MeV] \\ \hline 
 L      & 6    & 9.04 & 3.77  &  5.0  & -9.25  &  -0.28 & -0.27  & 117.0  \\ 
 Q      & 6    & 9.18 & 2.63 &  5.0  & -13.57 &  1.19  & -0.26  & 112.0  \\
 C      & 1.5  & 9.67 & -3.54 & -10.0 & -11.54 &  -2.88 & -0.07  & 114.0  \\
 C$_a$  &  0   & 9.02 &-12.33 & -20.0 & -7.96  &  -4.86 & 0.514 & 118.0   
   \et
\end{table}

\begin{thebibliography}{Aaa99a}
\bibitem{heid94} E.\ K.\ Heide, S.\ Rudaz, and P.\ J.\ Ellis,  
                        Nucl.\ Phys.\ {\bf A571}, 713 (1994)

\bibitem{cart95}  G.\ Carter, P.\ J.\ Ellis, and S.\ Rudaz,
                        Nucl.\ Phys.\ {\bf A603} 367 (1996);
                  Erratum-ibid. {\bf A608} 514 (1996) 

\bibitem{furn95} R.\ J.\ Furnstahl, H.\ B.\ Tang, and B.\ D.\ Serot, 
                       Phys.\ Rev.\ C {\bf 52} 1368 (1995)

\bibitem{mish93} I.\ Mishustin, J.\ Bondorf, and M.\ Rho,  
                        Nucl.\ Phys.\ {\bf A555} 215 (1993)

\bibitem{paper1} P.\ Papazoglou, J.\ Schaffner, S.\ Schramm,
                 D.\ Zschiesche, H.\ St\"ocker, and W.\ Greiner, 
                 Phys.\ Rev. C {\bf 55} 1499 (1997)  

\bibitem{strangenuc} T.\ Muto and T.\ Tatsumi, 
Phys.\ Lett. {\bf B283}, 165 (1992)

\bibitem{scha93} J.\ Schaffner, C.\ B.\ Dover, A.\ Gal, C.\ Greiner, 
       and H.\ St\"ocker,  Phys.\ Rev.\ Lett.\ {\bf 71}, 1328 (1993)

\bibitem{Eight64} M.\ Gell-Mann and Y.\ Neeman, 
             ``{\it The Eightfold Way}'', 
                W.\ A.\ Benjamin, Inc., New York and Amsterdam, 1964.
           
\bibitem{levy67} M.\ L\'evy, 
                    Nuovo Cimento {\bf 8}, 23 (1967)

\bibitem{gasi69} S. \ Gasiorowicz and D. \ Geffen, 
                 Rev. Mod. Phys. {\bf 41}, 531 (1969)

\bibitem{sche71}  J.\ Schechter, and Y. Ueda, 
                  Phys.\ Rev.\ D {\bf 3} 168 (1971). 

\bibitem{sche80} J.\ Schechter, 
                     Phys.\ Rev.\ D{\bf 21}, 3393 (1980)  

\bibitem{kerman} A.\ K.\ Kerman und  L.\ D.\ Miller,
        ``{\it Proceedings of the Second Relativistic Heavy Ion Summer 
        Study}'', 
             Berkeley, 1974

\bibitem{marsh65} R.\ E.\ Marshak, N.\ Mukunda, and S.\ Okubo, 
                  Phys.\ Rev.\ {\bf 137}, B699 (1965)

\bibitem{gell64}  M.\ Gell-Mann, 
                                 Physics,  {\bf 1}, 63 (1964)

\bibitem{ioff81} B.\ L.\ Ioffe, 
                        Nucl.\ Phys.\ {\bf B188}, 317 (1981)

\bibitem{chri86} G.\ A.\ Christos, 
                        Phys.\ Rev.\ D {\bf 35}, 330 (1987)

\bibitem{sche69} J.\ Schechter, Y. Ueda, and G.\ Venturi,  
                  Phys.\ Rev.\ {\bf 177} 2311 (1969) 

\bibitem{saku69}  J.\ J.\ Sakurai, 
             ``{\it Currents and Mesons}'', University of Chicago Press, 
                Chicago (1969)

\bibitem{cole85} S.\ Coleman, 
                  ``{\it Aspects of Symmetry}'', \\
                  Cambridge University Press; Cambridge, (1985)

\bibitem{mitt68} P.\ K.\ Mitter and L.\ J.\ Swank, 
                 Nucl.\ Phys.\ {\bf B8} 205 (1968)
                  
\bibitem{ko94} P.\ Ko and S.\ Rudaz, Phys.\ Rev.\ D {\bf 50} 6877 (1994)

\bibitem{sero86}  B.\ D.\ Serot and J.\ D.\ Walecka,
                  Adv.\ Nucl.\ Phys.\ {\bf 16}, 1 (1986) 

\bibitem{hugo58} N.\ M.\ Hugenholtz, L.\ van Hove, 
                       Physica 24, 363 (1958)
\bibitem{paper4} D.\ Zschiesche, P.\ Papazoglou, S.\ Schramm, 
       H.\ St\"ocker and W.\ Greiner, in preparation 

\bibitem{kuon95} H.\ Kuono, N.\ Kakuta, N.\ Noda, T.\ Mitsumori, 
                     A.\ Hasegawa,  
                    Phys.\ Rev.\ C {\bf 51}, 1754 (1995) 

\bibitem{sero97} B.\ D.\ Serot and J.\ D.\ Walecka, Report No., 
                 nucl-th/9701058

\bibitem{Dover84} C.\ B.\  Dover and A.\ Gal,  
                        Prog.\ Part.\  Nucl.\ Phys.\, {\bf 12}, 171 (1984)

\bibitem{ma95} J.\ Mares, E.\ Friedman, A.\ Gal and B.\ K.\ Jennings, 
                  Nucl. Phys. {\bf A594} (1995) 311    

\bibitem{Dov89a} C.\ B.\  Dover, D.\ J.\ Millener, and A.\ Gal,  
                        Phys.\ Rep.\  {\bf 184}, 1 (1989)

\bibitem{jennings90} B.\ K.\ Jennings, Phys.\ Lett.\ B {\bf 246}, 325 (1990)

\bibitem{bern68} J.\ Bernstein, 
             ``{\it Elementary Particles and Their Currents}'', \\
                Freeman and Company, San Francisco and London, 1968.

\bibitem{enhance}  P.\ J.\ de\ A.\ Bicudo, Phys.\ Rev.\ Lett.\ {\bf 72}, 
1600 (1994) 

\bibitem{paper3} P.\ Papazoglou, S.\ Schramm, H.\ St\"ocker, 
and W.\ Greiner, in preparation 

\bibitem{wein68} S.\ Weinberg, Phys.\ Rev.\ {\bf 166} 1568 (1968)

\bibitem{cole69} A.\ Coleman, J.\ Wess, and B.\ Zumino, Phys.\ Rev.\ {\bf 177} 
2239 (1969)

\bibitem{callen69} C.\ G.\ Callen, S.\ Coleman, J.\ Wess, and B.\ Zumino, Phys.\ 
Rev.\ {\bf 177}, 2247 (1969)

\bibitem{stoks96} V.\ J.\ G.\ Stoks and Th.\ A.\ Rijken, 
             Nucl.\ Phys.\ {\bf A613} 311 (1997)

\bibitem{bwlee69}  W.\ A.\ Bardeen and B.\ W.\ Lee
                  Phys.\ Rev.\ {\bf 177}, 2389 (1969)

\bibitem{thera2} P.\ Papazoglou, D.\ Zschiesche, S.\ Schramm, H.\ St\"ocker, 
                and W.\ Greiner, Report No. nucl-th/9706013,\\ {\it Proceedings 
               of the Int.\ Conference on Strangeness and Quark Matter}, 
               Santorini, to appear in J.\ Phys.\ G, 

\end{thebibliography}
\end{document}